\documentclass[prl,twocolumn]{revtex4}

\usepackage{latexsym,amsmath,graphics,graphicx}

\begin{document}
\title{Fe clusters on Ni and Cu: size and shape dependence of the spin moment}
\author{Ph.~Mavropoulos, S.~Lounis, R.~Zeller, and S.~Bl\"ugel}
\affiliation{Institut f\"ur
Festk\"orperforschung, Forschungszentrum J\"ulich, D-52425 J\"ulich,
Germany}
\date{\today}

\begin{abstract}
We present ab-initio calculations of the electronic structure of small
Fe clusters (1-9 atoms) on Ni(001), Ni(111), Cu(001) and Cu(111)
surfaces. Emphasis is given on the spin moments, and their dependence
on cluster size and shape. We derive a simple quantitative rule, which
relates the moment of each Fe atom linearly to its coordination
number. Thus, for an arbitrary Fe cluster the spin moment of the
cluster and of the individual Fe atoms can be readily found if the
positions of the atoms are known.
\end{abstract}
\maketitle

Small atomic clusters on surfaces constitute an extremely interesting
subject, as their electronic structure lies in the transition from the
behaviour in bulk to that of free molecules or atoms. In particular,
the electronic structure is characteristic of the cluster in that it
depends on the cluster atoms, shape, size, and orientation, as well as
on the substrate material on which the cluster is deposited. Thus it
is {\it terra incognita} for the electronic properties. Moreover,
since they are not hidden in the bulk, such structures are directly
accessible by a number of experimental techniques, such as scanning
tunneling microscopy, x-ray magnetic circular dichroism,
photoemission, etc.

In the area of magnetism, small clusters of magnetic atoms on surfaces
are expected to experience an enhancement of the magnetic
properties. As the cluster size decreases and the average coordination
of the atoms becomes smaller, the decreased hybridisation of the
atomic wavefunctions should lead to more pronounced magnetic
effects. Such trends with cluster size have been monitored in recent
experimental works; for example, using x-ray magnetic circular
dichroism, Lau and co-workers~\cite{Lau02} have focused on Fe/Ni(001)
systems, while Gambardella and collaborators \cite{Gambardella03} have
examined Co clusters on Pt(111); and nuclear resonant scattering of
synchrotron radiation has been employed for the study of Fe islands on
W(110)~\cite{Rohlsberger01}. Also other techniques, such as perturbed
angular correlation~\cite{Potzger02} or spin-polarized scanning
tunneling spectroscopy~\cite{Pietzsch04}, allow the analysis of
magnetic atoms on surfaces in the sub-monolayer regime, giving further
impetus to the field.

Similar to the ferromagnetic Fe, Co, and Ni clusters, V, Cr, and Mn
clusters should also have a strong intra-atomic exchange field; but,
in contrast to the ferromagnetic clusters, their inter-atomic exchange
should be antiferromagnetic, and due to competing interactions the
resulting magnetic order is in general non-collinear and depends on
individual details. Thus it will differ among clusters, and to
investigate the individual magnetic order of such clusters is a
challenge for both experiment and theory.

But even for homo-atomic and mass-selected clusters it is very
difficult to address experimentally the magnetic properties of each
one individually, let alone each atom in a cluster. In many cases just
the mean moment of clusters of a particular size or the average mean
magnetic moment per atom is found, averaged over an ensemble of
clusters of the same size but different shapes. This motivated us to
investigate the local and global spin moments of small Fe clusters (up
to 9 atoms) of monoatomic height and varying shape and size with a
material specific theory of predictive power, namely density
functional theory.  In this paper we summarise our results by
establishing a simple quantitative rule which relates the local spin
moment of an Fe atom in the cluster linearly to the coordination
number of its nearest neighbour atoms, with little influence of the
structure of the atoms in the rest of the cluster. In this way for any
arbitrary Fe cluster the spin moment of the cluster and of the
individual Fe atoms can be readily found if the positions of the atoms
are known. Even if the cluster shapes and sizes are not exactly known
experimentally, by making a reasonable assumption of the variation of
sizes and shapes the proposed rule allows an estimation of the
variation of the total moment across the deposited clusters.
Motivated by the work of Lau~{\it et al.}~\cite{Lau02} we concentrated
our attention on Fe clusters on the (001) and (111) surfaces of the
ferromagnetic substrate Ni and the non-magnetic substrate Cu.

First-principles calculations of clusters on surfaces can be
computationally quite demanding, considering that such systems break
the translational symmetry in all three directions. Most {\it
ab-initio} methods explicitly exploit the translational symmetry and
the clusters on surfaces are approximated by huge, computationally
expensive supercells which are periodically repeated. So far they have
been applied mostly for magnetic adatoms and chains~\cite{Spisak02}. A
breakthrough in the treatment of magnetic clusters was the development
of the Green function method of Korringa, Kohn and
Rostoker~\cite{Wildberger95,Papanikolaou02} for the embedding of
clusters at surfaces. Subsequent
applications~\cite{Stepanyuk99,Cabria02,Lazarovits02} established this
method as a powerful tool.

Our calculations are based on density-functional theory in the local
spin-density approximation~\cite{Vosko} (LSDA). The Green function
method of Korringa, Kohn, and Rostoker (KKR) is
employed~\cite{Papanikolaou02} to determine the spin density and
effective potential. Within the method, the Green function of the
perturbed system, {\it i.e.}, surface plus cluster, is related to the
one of the ``host'' reference system (the clean substrate) by a Dyson
equation. In this way the correct host boundary conditions are
included automatically in the Green function, thus no supercell
construction is needed. In order to allow for a screening of the
perturbation induced by the cluster, at least the first neighbouring
sites of all cluster atoms were considered in the self-consistency
cycle. The angular momentum expansion of the Green functions was
truncated at $l_\mathrm{max}=3$.  Tests have shown that these
parameters give reliable results for these systems.

A full-potential approach with the correct description of the atomic
cells~\cite{Stefanou90} was used. Since we are focusing here on trend
calculations, the atoms in the clusters were situated in the unrelaxed
lattice positions, using the LSDA equilibrium lattice parameter of
6.46~au (3.417~\AA ) for Ni and 6.63~au (3.507~\AA ) for
Cu~\cite{Asato99}.

\begin{figure}
\begin{center}
\includegraphics[width=0.9\linewidth]{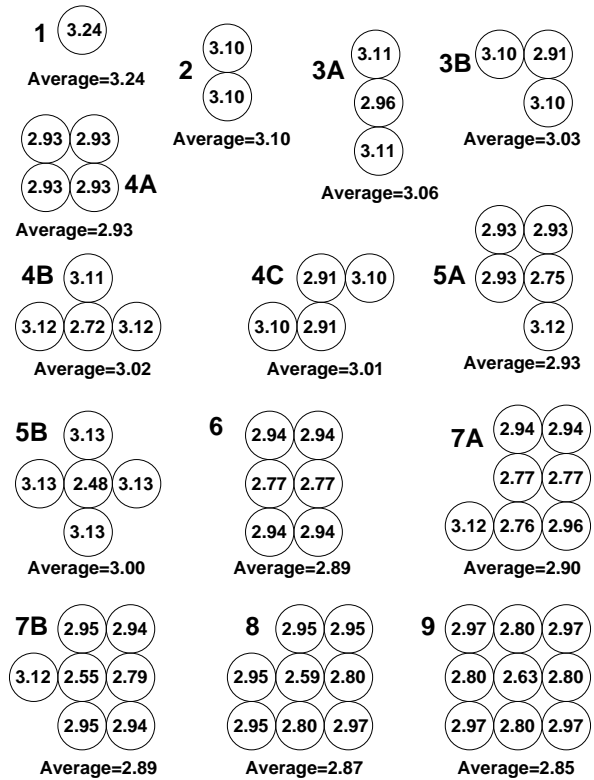}
\includegraphics[angle=270,width=0.9\linewidth]{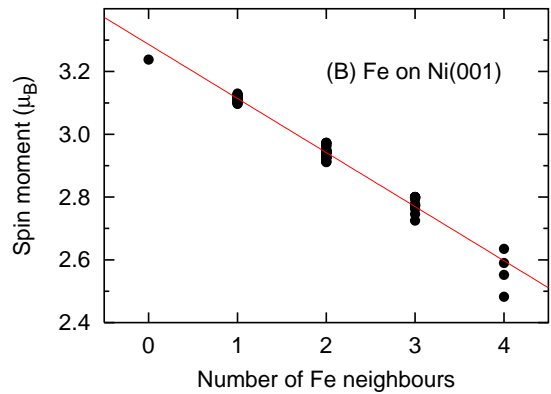}
\caption{A: Spin moment (in $\mu_B$) of atoms in 14 different Fe
clusters ranging from 1 to 9 atoms on the Ni(001) surface, and average
(per atom) moment of the clusters (the view is surface-adapted, {\it
i.e.}, rotated by 45$^\circ$ with respect to the in-plane fcc cubic
axes; the clusters are viewed from the top, {\it i.e.}, all atoms lie
on the surface). B: Linear trend for the atomic Fe spin moment as
function of the coordination number of nearest Fe neighbours.}
\label{Fig:1}
\end{center}
\end{figure}

The clusters considered on the Ni(001) surface are shown schematically
in Figure~\ref{Fig:1} viewed from the top (all atoms lie on the
surface). The view is adapted to surface geometry, meaning that it is
rotated by 45$^\circ$ with respect to the in-plane fcc cubic axes. The
smallest cluster is a single Fe adatom, while the largest consists of
9 Fe atoms. In each atom, the calculated spin moment is written, and
the average (per atom) moment of each cluster is also given. The Fe
moment is always ferromagnetically coupled to the Ni substrate moment
and to the Fe moments within the clusters.
Already at a first glance it is obvious that the average moment of the
clusters depends on the cluster size. The single adatom has manifestly
the highest moment (3.24~$\mu_B$), while the 9-atom cluster shows a
lower average moment of 2.85~$\mu_B$. 

This behaviour is expected on the grounds of hybridisation of the
atomic d levels with the neighbours. Atoms in larger clusters have, on
the average, higher coordination, thus their d wavefunctions are more
hybridised; this leads to lesser localisation and lesser tendency to
magnetism.

To pursue this idea further, we tried to correlate the local atomic
spin moment to the nearest neighbour coordination of each atom,
irrespective of the form or size of the cluster. For instance, let us
focus on all Fe atoms which have only one first Fe neighbour, {\it
i.e.}, $N_c=1$ (the coordination to the substrate is the same,
$N_s=4$, for all Fe atoms). Such atoms appear in the clusters with
size 2, 3, 4, 5, and 7; there are, in total, 10 such examples (having
excluded cases which are trivially equivalent by symmetry). \emph{All}
of them have spin moments ranging in the small interval between 3.10
and 3.13~$\mu_B$. Similarly, for the Fe atoms with two Fe neighbours
the spin moment ranges from 2.91 to 2.97~$\mu_B$. Collecting all
possible cases, from $N_c=0$ (single adatom) to $N_c=4$, we present
the results in Figure~\ref{Fig:1}B.

A surprising feature is the almost linear dependence of the spin
moment on the coordination number. In the general case, for arbitrary
magnetic atoms, this is not expected. For example,
Ref.~\onlinecite{Wildberger95} shows a study of the magnetism of small
4d atom clusters on Ag(001). There, it is reported that the
susceptibility is highly non-local, resulting even in an increase of
the spin moment with increasing coordination. This is related to the
larger extent of the 4d wavefunctions compared to the 3d ones of
Fe. In our case, Fe has a strong intra-atomic exchange field, arising
from rather localised 3d wavefunctions and their positioning with
respect to the Fermi level $E_F$. In a non-magnetic picture, $E_F$ is
well within the d virtual bound state, and the non-magnetic density of
states at $E_F$, $n(E_F)$, is always well above the transition point
to the magnetic state: the Stoner criterion $I\cdot n(E_F)>1$ (with
$I$ being the exchange integral) is safely fulfilled. Thus, the
hybridisation of the 3d levels with the neighbours' wavefunctions does
not affect the nonmagnetic $n(E_F)$ critically. With increasing
coordination, $n(E_F)$ is gradually lowered, and so is the
intra-atomic exchange field. The tiny variation of the local moments
for Fe atoms with the same nearest neighbour coordination but being in
different clusters may arise from intra-cluster interference effects
or indirectly via the different polarisation of the substrate.

Preliminary results on Co clusters, for which the 3d virtual bound
state is lower in energy, show that this linear behaviour holds to a
lesser extent, and much less so for Ni, where the intra-atomic
exchange field is even weaker~\cite{Elsewhere}. A similar difference
among Fe, Co, and Ni clusters on Ag(001) has been shown in
Ref.~\cite{Lazarovits02} and~\cite{Stepanyuk99}.

On the other hand, Mn and Cr should have a strong intra-atomic
exchange field, but they present tendency for antiferromagnetism or
even non-collinear magnetic structures~\cite{Elsewhere}. In this
respect, Fe is expected to be perhaps unique among the transition
elements in showing such a clear cut linear trend.

\begin{figure}
\begin{center}
\includegraphics[width=0.9\linewidth]{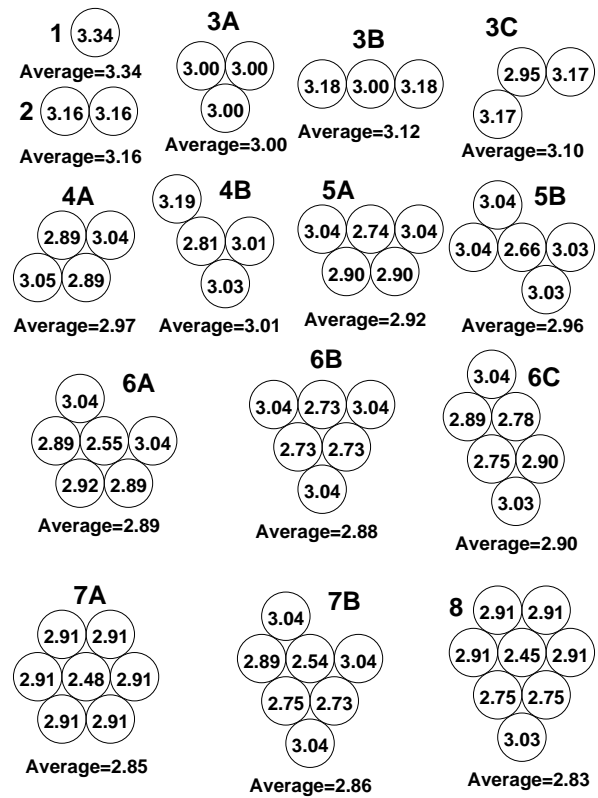}
\includegraphics[angle=270,width=0.9\linewidth]{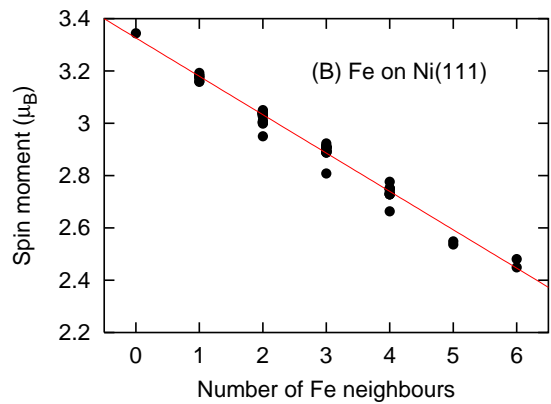}
\caption{A: Spin moments (in $\mu_B$) of atoms in 15 different Fe
clusters ranging from 1 to 8 atoms on the Ni(111) surface, and average
(per atom) moments of the clusters (the clusters are viewed from the
top, {\it i.e.}, all atoms lie on the surface). B: Linear trend for the
atomic Fe spin moment as function of the coordination number of Fe
neighbours.}
\label{Fig:2}
\end{center}
\end{figure}

Similar is the case of Fe clusters on the more compact Ni(111)
substrate. The considered clusters, sized up to 8 atoms, are shown
together with the spin moments in Figure~\ref{Fig:2}A. Since the Fe
adatom on the (111) surface has only 3 Ni neighbours instead of 4 on
the (001) surface, its spin moment is higher than on the (001)
surface. On the other hand, the number of Fe neighbours (in-plane) can
increase up to $N_c=6$, which makes the Fe moment decrease up to
2.45~$\mu_B$; the collected statistics (spin moments {\it vs.}~$N_c$)
are presented in Figure~\ref{Fig:2}B.
Again we see a linear behaviour of the moment {\it vs.}~$N_c$.

Next we turn to Fe clusters on Cu(001) and Cu(111) surfaces, for which
we considered the same types of clusters as in the case of Ni. Since
the Cu lattice parameter is slightly larger than the one of Ni, one
might expect a lesser degree of hybridisation and therefore increased
spin moments at the Fe atoms. Nevertheless, the moments are slightly
lower, presumably because in the case of Ni the magnetisation of the
substrate assists the spin polarisation of the cluster.
\begin{figure}
\begin{center}
\includegraphics[angle=270,width=0.9\linewidth]{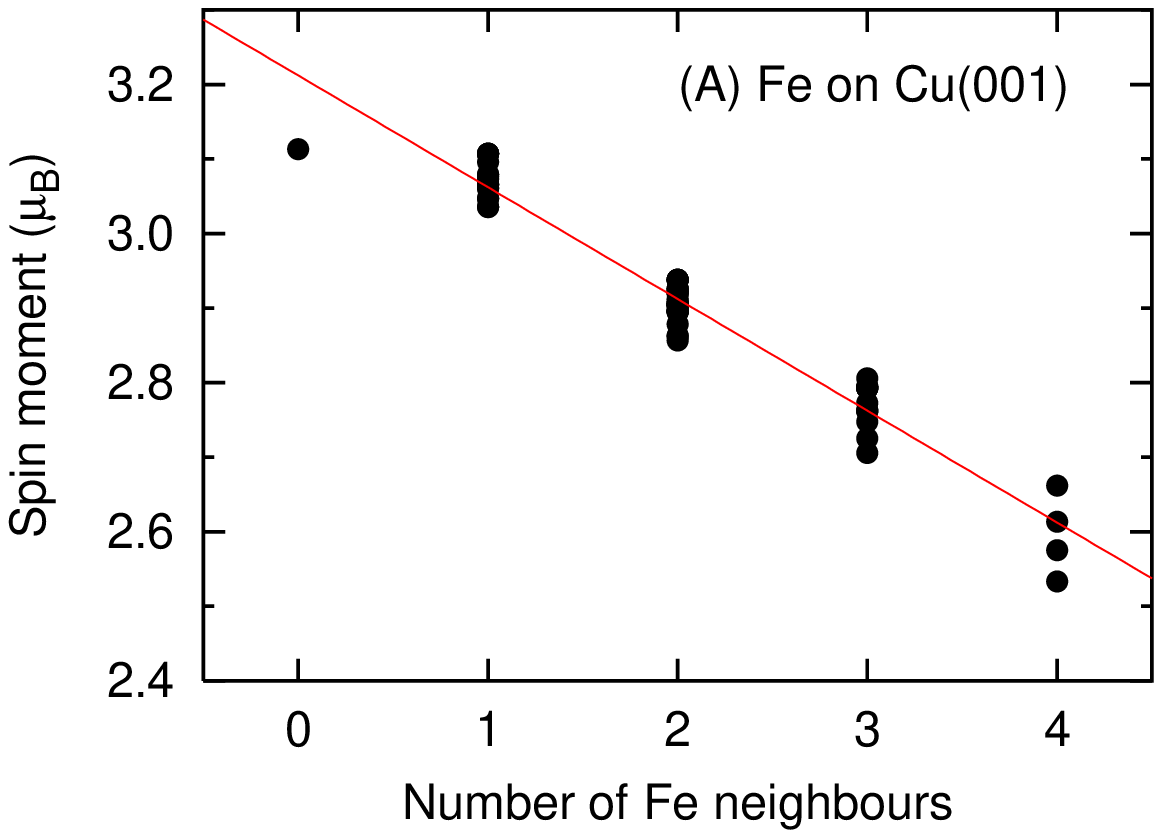}
\includegraphics[angle=270,width=0.9\linewidth]{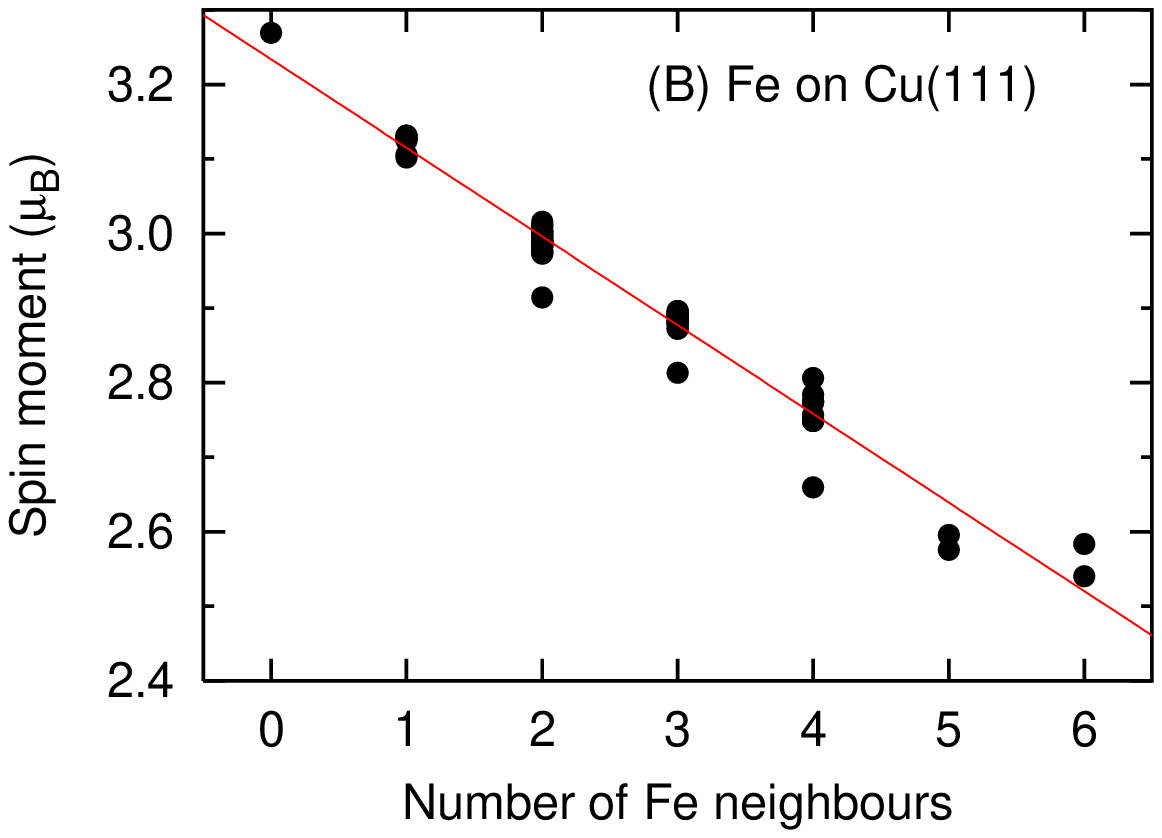}
\caption{Linear trend for the atomic Fe spin moment as function of the
coordination in Fe clusters on Cu surfaces. A: on Cu(001), and B: on Cu(111).}
\label{Fig:3}
\end{center}
\end{figure}
The results for clusters on Cu(001) and on Cu(111) are shown in
Figure~\ref{Fig:3}A and ~\ref{Fig:3}B respectively. The result for the
single adatom, showing again the highest moment, should not be taken
literally: it is well known that magnetic impurity atoms in the bulk
and on the surface of nonmagnetic hosts enter the Kondo regime at low
temperatures, whence the spin moment shows quantum fluctuations and
has a net average of zero. This regime is not accessible by a mean
field theory as the LSDA. Thus the result shown here for the adatom
should be considered either as the trend of the LSDA calculation, or
what one expects above the Kondo temperature. For magnetic clusters
(dimers and larger) the quantum fluctuations of the moment are
strongly suppressed and the Kondo temperature decreases drastically
with cluster size; thus the results given here for the clusters are
realistic.

The linear dependence of the spin moments is again evident, except
perhaps the result for the single adatom ($N_c=0$).  A fit of the
calculated data, using a linear equation of the form
\begin{equation}
M=-aN_c+b,
\end{equation}
where $M$ is the local spin moment, results in values for $a$ and $b$
given in Table~\ref{Tab:1}. We note that calculations on Fe
nanostructures on Ag(001), reported for instance in
Refs.~\cite{Stepanyuk99} and \cite{Lazarovits02} (using the KKR
method), and~\cite{Izquierdo00} (using the SIESTA code), show a
similar trend, although in these works fewer kinds of clusters are
presented.
\begin{table}
\begin{center}
\begin{tabular}{ccc}
\hline
Surf.   & $a(\mu_B)$   & $b(\mu_B)$  \\
\hline
Ni(001) & 0.17  & 3.29  \\
Ni(111) & 0.15  & 3.33  \\
Cu(001) & 0.15  & 3.21  \\
Cu(111) & 0.12  & 3.23  \\
\hline
\end{tabular}
\caption{Parameters $a$ and $b$ from the linear fit of the spin moment
  according to Eq.~(1).}
\label{Tab:1}
\end{center}
\end{table}

In summary, we have presented first-principles calculations of small
Fe clusters on Ni and Cu (001) and (111) surfaces. The average (per
atom) spin moments of the clusters are reduced with cluster size,
mainly because the higher average coordination of the Fe atoms
increases the hybridisation of the d states. The important parameter
turns out to be the local coordination of each distinct Fe atom,
irrespectively of the exact size or shape of the cluster. A linear
relation of the coordination to the atomic spin moment has been found
in a good approximation.

\begin{acknowledgments}
We would like to thank P.~H.~Dederichs for fruitful discussions.
Financial support by the Priority Programme ``Clusters in Contact with
Surfaces'' (SPP1153) of the Deutsche Forschungsgemeinschaft is
gratefully acknowledged.
\end{acknowledgments}



\begin{thebibliography}{99}
\bibitem{Elsewhere} Calculations for such systems are in progress.
Detailed results and conclusions will be presented elsewhere.
\end{thebibliography}
\end{document}